\documentclass[10pt,a4paper,twocolumn,superscriptaddress,prd,eqsecnum]{revtex4}
\usepackage[english]{babel}
\usepackage{amsmath}
\usepackage{amsfonts}
\usepackage{amssymb}
\usepackage{bm}
\usepackage{tensor}

\renewcommand{\d}{\mathop{}\!\mathrm{d}}
\newcommand{\dvinv}{\, \sqrt{-g} \, \mathop{}\!\mathrm{d}^4\!}

\newcommand{\partd}[2][]{\frac{\partial #1}{\partial #2}}

\newcommand{\D}[1][]{\nabla_{\!#1}}

\newcommand{\vev}[1]{\big< #1 \big>}

\newcommand{\zmag}{z_{\scriptscriptstyle >}}
\newcommand{\zmin}{z_{\scriptscriptstyle <}}
\newcommand{\col}[2]
{\begin{pmatrix}
												#1\\												#2
										\end{pmatrix}}

\begin{document}

\title{Energy-momentum tensor for a \textit{scalar} Casimir apparatus in 
a weak gravitational field: Neumann conditions}

\author{George M. Napolitano}
\affiliation{Dipartimento di Scienze Fisiche, Complesso 
Universitario di Monte S. Angelo, Via Cintia, Edificio 6, 80126 Napoli, Italy}

\author{Giampiero Esposito}
\affiliation{Istituto Nazionale di Fisica Nucleare, Sezione di Napoli, 
Complesso Universitario di Monte S. Angelo, Via Cintia, Edificio 6, 
80126 Napoli, Italy}

\author{Luigi Rosa}
\affiliation{Istituto Nazionale di Fisica Nucleare, Sezione di Napoli, 
Complesso Universitario di Monte S. Angelo, Via Cintia, 
Edificio 6, 80126 Napoli, Italy}
\affiliation{Dipartimento di Scienze Fisiche, Complesso 
Universitario di Monte S. Angelo, Via Cintia, Edificio 6, 80126 Napoli, Italy}

\begin{abstract}
We consider a Casimir apparatus consisting of two perfectly 
conducting parallel plates, subject to the weak gravitational 
field of the Earth. The aim of this paper is the calculation of 
the energy-momentum tensor of this system for a free, real massless 
scalar field satisfying Neumann boundary conditions on the plates. 
The small gravity acceleration (here considered as not varying between 
the two plates) allows us to perform all calculations to  
first order in this parameter. Some interesting results are found: 
a correction, depending on the gravity acceleration, to the well-known 
Casimir energy and pressure on the plates. Moreover, this scheme predicts 
a tiny force in the upwards direction acting on the apparatus. These 
results are supported by two consistency checks: 
the covariant conservation of the energy-momentum tensor and the 
vanishing of its regularized trace, when the scalar field is conformally 
coupled to gravity. 
\end{abstract}

\maketitle

\section{Introduction}
Quantum field theory in curved spacetime, although far from being  
a definitive theory unifying the quantum theories with gravitation, 
offers nevertheless some intriguing results, such as the well-known 
Hawking radiation \cite{b:Hawk} and the closely related Unruh 
effect \cite{b:Unruh}.
Moreover, in the recent literature, a number of papers studying 
the influence of a gravitational field on the energy stored in a 
Casimir cavity appeared \cite{b:Sorge, 025004, prd78BER, b:prd77ENR}, 
in particular nowadays the theoretical prediction that the vacuum 
fluctuations follow the equivalence principle seems to be 
demonstrated \cite{025004,prd78BER,2564,MPWSRF08}.

The main result is that there seems to be full agreement on the fact that  
Casimir energy gravitates, i.e. a Casimir cavity storing an energy 
$E_0$ experiences a force of magnitude $F= \frac{\mathfrak{g}}{c^2}|E_0|$, 
$\mathfrak{g}$ being the gravity acceleration.
The present work is the natural development of our previous 
article \cite{b:prd77ENR}, where we computed the vacuum expectation 
value of the renormalized energy-momentum tensor of a massless 
scalar field in a Casimir cavity. The scalar field was there assumed to 
satisfy Dirichlet boundary conditions on the parallel plates constituting 
the cavity. The analogy with the electromagnetic case \cite{prd74BCER}, 
where the components of the potential satisfy a mixture 
of Dirichlet and Neumann conditions, motivated the 
present analysis. Here we show that the Neumann 
boundary conditions yield equivalent results. Interestingly, combining 
these results with those obtained in \cite{b:prd77ENR}, i.e. considering a 
two-component field satisfying mixed boundary condition, the electromagnetic 
case \cite{prd74BCER,prd78BER} is exactly reproduced.

As expected, the conformal and minimal coupling of the scalar field with 
gravity yield different results. We find that the known divergences 
on the boundaries \cite{b:DandC} (the plates of the apparatus) lead to 
finite physical quantities only in the conformal coupling case. Quite 
different is the mixed boundary conditions case considered in the second 
part of Sec. III, where we find that the energy stored and the 
pressures are independent of the coupling constant $\xi$, to first
order in the gravity acceleration. 

\section{Energy-momentum tensor}
\label{sec:CompEMT}

Since we rely heavily on our work in Ref. \cite{b:prd77ENR}, we refer
the reader to it for all technical details. It is enough to say that,
starting from the basic formalism for scalar fields in curved
spacetime \cite{b:Misn, b:Birr}, we use the covariant geodesic point
separation method of Ref. \cite{b:Chris76} to expand the Green functions
to first order in the parameter $\epsilon \equiv {2 \mathfrak{g} a
\over c^{2}}$, $a$ being the distance between the plates. 
By virtue of translation invariance, one can perform a
Fourier analysis of the Green functions, with the associated {\it reduced}
Green functions, which obey Neumann boundary conditions on parallel
plates, i.e. 
\begin{equation}
\partd[\gamma^{(i)}]{z}\bigg|_{z=0} 
= \partd[\gamma^{(i)}]{z}\bigg|_{z=a} = 0. \qquad i = 0,1.
\label{eq:NeuCond2}
\end{equation}
We therefore obtain, to zeroth order in $\epsilon$,
\begin{equation}
\gamma^{(0)}(z,z') = - \frac{\cos (\lambda\zmin) \, 
\cos (\lambda (a-\zmag))}{\lambda \, \sin (\lambda a)},
\label{eq:gamma0}
\end{equation}
where $\zmin \equiv \min\{z,z'\}$ and $\zmag \equiv \max\{z,z'\}$, and,
to first order in $\epsilon$,
\begin{equation}
\begin{split}
\gamma^{(1)}(z,z') = & \frac{1}{4 a \lambda^2} 
\bigg\{ \bigg[ (k_0^2-\lambda^2) (z+z') \\
&- k_0^2 \left( z^2 \partd{z} + z'^2 \partd{z'}\right) \\
& + \left( \frac{k_0^2}{\lambda^2} - 1 \right) 
\left(\partd{z} + \partd{z'}\right) \bigg] \gamma^{(0)}(z,z') \\
&- a^2 k_0^2 \frac{\cos (\lambda z) \, 
\cos (\lambda z')}{\sin^2 (\lambda a)} \bigg\}.
\end{split}
\label{eq:gamma1}
\end{equation}
Renormalization of the energy-momentum tensor in curved spacetime is 
often carried out by subtracting the stress tensor constructed by the 
Schwinger--DeWitt method \cite{b:Chris76, PRPLC}. Here we follow the 
classical scheme of renormalization of the Casimir effect in flat space, 
i.e we subtract, from the energy-momentum tensor of the system, the one 
associated to a field propagating in free space (without boundaries). 
Therefore we need also the free Feynman Green functions. After 
having obtained the full Feynman Green functions, the Hadamard functions
are twice the imaginary part of the Feynman, and 
we can evaluate the energy-momentum tensor up to first order of the expansion
\begin{equation}
\vev{T_{\mu\nu}} \sim \vev{T^{(0)}_{\mu\nu}} 
+ \epsilon \vev{T^{(1)}_{\mu\nu}} + O(\epsilon^2).
\end{equation}

On defining the new variables $s \equiv \frac{\pi z}{a}$ 
and $s'\equiv\frac{\pi z'}{a}$, the renormalized energy-momentum tensor, 
to zeroth order in $\epsilon$, is
\begin{equation}
\begin{split}
\vev{T^{(0)}_{\mu\nu}} & = -\frac{\pi^2}{1440a^4}
{\rm diag}(1,-1,-1,3) \\
&-\left( \xi - \frac{1}{6} \right) \frac{2 + \cos 2s}{8 a^4 \sin^4 s}
{\rm diag}(1,-1,-1,0),
\end{split}
\end{equation}
while, to first order, the only nonvanishing components are found to be 
\begin{equation}
\begin{split}
\vev{T^{(1)}_{00}} = & \frac{\pi \csc^2 s}{14400a^4} 
\left\{ 77\pi + 141s - 20[7+2(\pi-s)s] \cot s \right.\\ 
&\left.+ 30 [-4\pi+3s+4(\pi-s)s \cot s] \csc^2 s 
+ \cos 2s \right.\\ 
&\left.\times (3\pi-s+150s \csc^2 s) \right\} 
+ \left( \xi - \frac{1}{6} \right) \frac{\pi \csc^2 s}{48a^4} \\
&\times \left\{ 4\pi-2s-\cot s [1+2(\pi-s)s+6s\cot s] \right.\\
&\left. + 6[-\pi+(\pi-s)s\cot s] \csc^2 s \right\},
\end{split} 
\label{eq:t100}
\end{equation}
\begin{equation}
\begin{split}
\vev{T^{(1)}_{11}} = & \frac{\pi \csc^5 s}{115200a^4} 
\left\{ 20 [-7+22(\pi-s)s]\cos s +20 \right.\\ 
&\left. \times [7+2(\pi-s)s]\cos3s 
+ (\pi-2s)(-230\sin s \right.\\
&\left. - 85 \sin 3s + \sin 5s) \right\} - \left( \xi 
- \frac{1}{6} \right) \frac{\pi\csc^5 s}{96a^4}\\
&\times \left\{ [1+11(\pi-s)s]\cos s + [-1+(\pi-s)s] \right.\\
&\left. \times \cos 3s - 2 (\pi-2s) (3\sin s + \sin3s)\right\}, 
\end{split}
\label{eq:t111}
\end{equation}
\begin{equation}
\vev{T^{(1)}_{22}} = \vev{T^{(1)}_{11}},
\label{eq:t122}
\end{equation}
\begin{equation}
\begin{split}
\vev{T^{(1)}_{33}} =  - \frac{\pi^2}{1440a^4} 
+ \frac{\pi s}{720a^4} - \left(\xi - \frac{1}{6} \right) 
\frac{\pi}{16a^4} \frac{\cos s}{\sin^3 s}.
\end{split}
\label{eq:t133}
\end{equation}

The consistency of this result is ensured by the following tests. 
First of all, the computed tensor is found to be covariantly conserved up 
to first order in the $\epsilon$ parameter, i.e. it satisfies the equation
$\D^\mu \vev{T_{\mu\nu}} = 0$.
On the other hand we know that, for a conformal scalar field, the following 
relation between the trace of the tensor and the mass of the field holds:
$ T_\mu^\mu = -2m^2 \phi^{2}$.
Hence we expect a vanishing trace when $\xi=\frac{1}{6}$, our scalar 
field being massless. This is exactly what we have found, 
because, upon defining
$\tau_\xi \equiv g^{\mu\nu} \vev{T_{\mu\nu}}$, 
we have
\begin{equation}
\begin{split}
\tau_\xi & = \left( \xi - \frac{1}{6} \right) 
\frac{\pi \csc^5s}{32a^4} \big\{ 6\pi (3 \sin s + \sin 3s) \\
&- \epsilon [(1 + 11(\pi - s)s) \cos s - (1 - (\pi - s)s) \cos 3s \\
&- 2 ( \pi - 2s ) (3 \sin s + \sin 3s) ] \big\},
\end{split}
\label{eq:trace}
\end{equation}
that clearly vanishes in the case of conformal coupling.

\section{Casimir energy and pressure}
\label{sec:EandP}
The energy density $\rho$ stored in our Casimir apparatus can be 
obtained by projecting the renormalized energy-momentum tensor along a unit 
timelike vector with components $u^\mu = \left(- 
\frac{1}{\sqrt{-g_{00}}},0,0,0 \right)$, so that
\begin{widetext}
\begin{equation}
\begin{split}
\rho &= \vev{T_{\mu\nu}} u^\mu u^\nu \\
& = - \frac{\pi^2}{1440a^4} + \epsilon \frac{\pi \csc^2 s}
{14400a^4} \bigg\{ 77 \pi + 146 s - 20 \big[7 + 2(\pi - s) s \big] 
\cot s + 30 \big[ -4 \pi + 3 s + 4 (\pi - s) s \cot s \big] \csc^2 s \\
&+ 3 \cos 2s (\pi - 2 s + 50 s \csc^2 s) \bigg\} 
+ \left( \xi - \frac{1}{6} \right) \bigg\{ 
- \frac{\pi^2 (2 + \cos 2s)}{8a^4 \sin^4 s} 
+ \epsilon \frac{\pi \csc^5 s}{192 a^4} \bigg[ (-1 
+ 22 (\pi - s) s) \cos s \\
&+ (1 + 2 (\pi - s) s) \cos 3s - 4 (\pi - 2 s) 
(3 \sin s + \sin 3s) \bigg] \bigg\} .
\end{split}
\end{equation}
\end{widetext}
Therefore the energy stored, following Ref. \cite{b:prd77ENR}, is
\begin{equation}
E = \frac{A a}{\pi} \lim_{\zeta \rightarrow 0^+} 
\int_\zeta^{\pi-\zeta} \d s \, \sqrt{-g} \, \rho,
\end{equation}
where $A$ is tha area of the plates. This yields
\begin{equation}
\begin{split}
E_\xi = &- \frac{\pi^2 A}{1440a^3} - \frac{\pi^2 A \epsilon}{5760a^3} 
- \left( \xi - \frac{1}{6} \right) \frac{\pi A}{4 a^3} \\ 
&\times \bigg( 1 + \frac{\epsilon}{4} \bigg) 
\lim_{\zeta\rightarrow 0^+} \frac{\cos\zeta}{\sin^3\zeta}.
\end{split}
\end{equation}
The conformal coupling case ($\xi = \frac{1}{6}$) ensures the finiteness 
of the above result that, reintroducing the constants $\hbar$, $c$ and 
the explicit expression of $\epsilon$, reads 
\begin{equation}
E_c = - \frac{\pi^2 \hbar c }{1440} \frac{A}{a^3} 
\left( 1 + \frac{1}{2} \frac{\mathfrak{g} a}{c^2} \right).
\end{equation}
With analogous arguments we find the pressure on the plates
\begin{equation}
P_\xi(z=0) = \frac{\pi^2}{480a^4} + \frac{\pi^2 \epsilon}{1440a^4} 
+ \left( \xi - \frac{1}{6} \right) \frac{\pi \epsilon}{16a^4} 
\lim_{s \rightarrow 0^+} \frac{\cos s}{\sin^3 s}
\end{equation}
and
\begin{equation}
P_\xi(z=a) = - \frac{\pi^2}{480a^4} + \frac{\pi^2 \epsilon}{1440a^4} 
- \left( \xi - \frac{1}{6} \right) \frac{\pi \epsilon}{16a^4} 
\lim_{s \rightarrow \pi^-} \frac{\cos s}{\sin^3 s}.
\end{equation}
Once again the divergent terms vanish when $\xi=\frac{1}{6}$, giving
\begin{align}
&P_c (z=0) = \frac{\pi^2}{480} \frac{\hbar c}{a^4} \left( 1 
+ \frac{2}{3} \frac{\mathfrak{g} a}{c^2} \right), \\
&P_c (z=a) = - \frac{\pi^2}{480} \frac{\hbar c}{a^4} 
\left( 1 - \frac{2}{3} \frac{\mathfrak{g} a}{c^2} \right).
\end{align}

The force acting on the system has to be calculated by considering 
the redshift $r$ of the point $\bar{z}$ where the pressures act, relative 
to the point $z_s$ where they are added, i.e. \cite{025008}
\begin{equation}
r(\bar{z},z_s) = \sqrt{\frac{|g_{00}(\bar{z})|}{|g_{00}(z_s)|}} 
\simeq 1 + \frac{\mathfrak{g}}{c^2} (\bar{z}- z_s).
\end{equation}
Thus, the net force obtained has magnitude
\begin{equation}
\begin{split}
F & = A [ P_c(0) \, r(0,z_s) + P_c(a) \, r(a,z_s) ]\\
& = \frac{\pi^2}{1440} \frac{A\hbar \mathfrak{g}}{c a^3} 
= \frac{\mathfrak{g}}{c^2} |E_0|,
\end{split}
\end{equation}
where we have defined $E_0 \equiv - 
\frac{\pi^2 \hbar c}{1440} \frac{A}{a^3}$. 
Therefore, direction (upwards along the $z$ axis) and magnitude of this force 
are in full agreement with the equivalence principle.

Note that some interesting effects result from combining 
the formulas here obtained with those in our previous 
work \cite{b:prd77ENR}, where the same problem with Dirichlet boundary 
conditions was considered. 
We start by defining the real massless two-component field
$\Phi = \col{\phi_D}{\phi_N}$,
where the subscripts $D$ and $N$ indicate that the components satisfy 
homogeneus Dirichlet and Neumann boundary conditions, respectively.
This is more than a toy model, because the electromagnetic Casimir 
effect with perfect-conductor boundary conditions on parallel plates
leads exactly to such a mixture of boundary conditions 
(see, for example, section 4.5 of Ref. \cite{Kluwer97}).

It is easy to see that, starting from the action functional 
$\mathcal{S}_t = -\frac{1}{2} \int (\Phi^\dag_{;\mu} \Phi^{;\mu} 
+ \xi R \Phi^\dag \Phi) \dvinv x$,
the vacuum expectation value of the renormalized energy-momentum 
tensor associated with this action reads
\begin{equation}
\vev{\mathcal{T}_{\mu\nu}} = \vev{T^D_{\mu\nu}} + \vev{T^N_{\mu\nu}},
\end{equation}
with obvious notation.

Thus, on combining the results of the Dirichlet and Neumann cases we have
\begin{equation}
\vev{\mathcal{T}^{(0)}_{\mu\nu}} =  -\frac{\pi^2}{720a^4}
{\rm diag}(1,-1,-1,3),
\end{equation}
and
\begin{equation}
\vev{\mathcal{T}^{(1)}_{00}} = - \frac{\pi^2}{1200a^4} 
\left( 1 - \frac{s}{3\pi} \right) + \left( \xi - \frac{1}{5} \right) 
\frac{\pi}{12a^4} \frac{\cos s}{\sin^3 s},
\end{equation}
\begin{equation}
\vev{\mathcal{T}^{(1)}_{11}} =  \frac{\pi^2}{3600a^4} 
\left( 1 - \frac{2s}{\pi} \right) - \left( \xi - \frac{3}{20} 
\right) \frac{\pi}{12a^4} \frac{\cos s}{\sin^3 s},
\end{equation}
\begin{equation}
\vev{\mathcal{T}^{(1)}_{22}} = \vev{\mathcal{T}^{(1)}_{11}},
\end{equation}
\begin{equation}
\vev{\mathcal{T}^{(1)}_{33}} = - \frac{\pi^2}{720a^4} 
\left( 1 - \frac{2s}{\pi} \right).
\end{equation}

Clearly, the covariant conservation holds, being satisfied 
separately for $\vev{T^D_{\mu\nu}}$ and $\vev{T^N_{\mu\nu}}$. The trace, 
defined in (\ref{eq:trace}), is found to be
\begin{equation}
\tau_t = - \left ( \xi - \frac{1}{6} \right) \frac{\pi \epsilon}{a^4} 
\frac{\cos s}{\sin^3 s},
\end{equation}
that once again is vanishing for $\xi=\frac{1}{6}$. 

The Casimir energy is twice the value in (3.4),
and the pressures on the plates are twice the values in (3.7) and (3.8).
Interestingly, in the mixed case, energy and pressure are 
both finite quantities for any value of $\xi$, to first order 
in $\epsilon$. Moreover, these results 
coincide perfectly with those found in the electromagnetic case 
\cite{prd74BCER}.

\section{Concluding remarks}

We have evaluated the vacuum expectation value 
of the renormalized energy-momentum tensor of a massless scalar field, 
satisfying Neumann boundary conditions on the parallel plates of a Casimir 
cavity immersed in a weak gravitational field. The calculations have been 
performed up to first order of the expansion in the parameter 
$\epsilon \equiv \frac{2\mathfrak{g}a}{c^2}$.

In agreement with the results found in \cite{prd74BCER} and with our 
previous work \cite{b:prd77ENR}, we have found a small correction to the 
Casimir energy not affected by the gravitational field, 
and the theoretical prediction that the whole 
cavity experiences a force proportional to the energy stored 
and with magnitude $F = \frac{\mathfrak{g}}{c^2} E_0$ 
(up to first order in $\mathfrak{g}$), 
pointing in the upwards direction. This result is in accordance with those 
found in \cite{025004, 2564, MPWSRF08} 
and seems to imply that Casimir energy gravitates, 
i.e. {\it the vacuum energy stored in a Casimir apparatus  
behaves, on theoretical ground, 
like a negative mass in a gravitational field}. 

The finiteness of the physical quantities was ensured by setting the 
coupling constant $\xi$ to $\frac{1}{6}$, i.e. only for conformal coupling 
between the scalar field and gravity, at least up to the order of our 
approximation; further investigations are needed to go to higher orders. 

A quite different situation appeared on considering a two-component field 
satisfying mixed boundary conditions. In this case, even though divergent 
boundary terms affect the energy-momentum tensor, this yields finite energy 
and pressures, in complete accordance with those found in \cite{prd74BCER}.
Other valuable related work can be found in Refs. \cite{S1,S2,S3,S4}, which 
focus on quantum field theory in Rindler spacetime. This point of view
has been later exploited in Refs. 
\cite{025004, prd78BER}, whose first-order results
agree with ours as we said before.

At this stage, further studies are required at least in two directions. First, 
we might try to use the technique here shown for the calculation of 
$\vev{T_{\mu\nu}}$ to the second (and even higher) 
order of the expansion in $\epsilon$, to further check
the agreement with the analysis of Ref. \cite{prd78BER}, which relies
instead on the uniform asymptotics of Bessel functions; moreover, the 
agreement of results obtained from 
different approaches persuades us to extend 
this kind of analysis to Casimir devices in other configurations \cite{ESA}.

\acknowledgments
G. Esposito is grateful to the Dipartimento di Scienze Fisiche of
Federico II University, Naples, for hospitality and support.

\bibliography{cas}

\end{document}